\pdfoutput=1
\documentclass[11pt,a4paper,twoside,openright,closeany,textany]{book}
\usepackage{makeidx}
\usepackage{graphicx}
\usepackage{paralist}
\usepackage{caption} 
\usepackage{amsmath}
\usepackage{amssymb}
\usepackage{mathtools}
\usepackage{fancyhdr}
\usepackage{multicol}
\usepackage{multirow}

\usepackage{mdframed}
\usepackage[usenames,dvipsnames,svgnames,table]{xcolor}
\usepackage{array}
\usepackage[%
  colorlinks=true,%
  linkcolor=black,%
  urlcolor=black,%
  citecolor=black%
]{hyperref}
\usepackage{nameref} 
\usepackage{ifthen} 
\usepackage[font={small,sl}]{caption}
\usepackage[authoryear]{natbib}
\renewcommand\bibname{References}
\newcommand{\mychapbib}{
  \addcontentsline{toc}{section}{\bibname}
  \bibliographystyle{natbib}
  \bibliography{strucbioinf}
}
\usepackage[colorinlistoftodos]{todonotes}
\usepackage{letltxmacro}

\usepackage{tocstyle} 
\usetocstyle{standard}
\settocfeature{raggedhook}{\raggedright}
\newcommand{\bibfont}{\small\raggedright}
\def\cite{\citep}

\LetLtxMacro{\oldTodo}{\todo}
\renewcommand{\todo}[2][]{\oldTodo[#1]{TODO: #2}}

\usepackage[normalem]{ulem}

\newcommand\inwish[1]{\oldTodo[inline,color=SkyBlue]{WISH: #1}}

\newboolean{onechapter}

\newcommand{\AF}[1][~]{K.\@#1Anton#1Feenstra}
\newcommand{\SA}[1][~]{Sanne#1Abeln}

\newcommand{\HM}[1][~]{Halima#1Mouhib}

\newcommand{\JvG}[1][~]{Juami#1H.\@#1M.\@#1van#1Gils}

\newcommand{\MD}[1][~]{Maurits#1Dijkstra}

\newcommand{\EvD}[1][~]{Erik#1van#1Dijk}

\newcommand{\AG}[1][~]{\mbox{Arthur}#1\mbox{Goetzee}}

\newcommand{\IH}[1][~]{\mbox{Isabel}#1\mbox{Houtkamp}}

\newcommand{\orcid}[1]{\href{https://orcid.org/#1}{\raisebox{-0.7ex}{\protect\includegraphics[height=3ex]{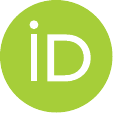}}}}
\definecolor{idgreen}{RGB}{166 206 57}
\newcommand{\mailid}[1]{\href{mailto:#1}{\raisebox{-0.3ex}{\color{idgreen}\textsf{\textbf{\Large \protect@}}}}}

\newcommand{\AFid}{\orcid{0000-0001-6755-9667}}
\newcommand{\SAid}{\orcid{0000-0002-2779-7174}}
\newcommand{\HMid}{\orcid{0000-0001-5031-3468}}

\newcommand{\JvGid}{\orcid{0000-0003-3706-7818}}

\newcommand{\AGid}{\orcid{0000-0001-7532-2627}}

\newcommand{\EvDid}{\orcid{0000-0002-6272-2039}}

\newcommand{\MDid}{\orcid{0000-0002-7971-6209}}

\newcommand{\IHid}{\orcid{0000-0002-4222-7292}}

\newcommand{\ACtxt}{Wrote the text}
\newcommand{\ACfig}{Created figures}
\newcommand{\ACref}{Review of current literature}
\newcommand{\ACeds}{Editorial responsibility}
\newcommand{\ACproof}{Critical proofreading}
\newcommand{\ACfb}{Non-expert feedback}

\newcommand{\Angs}[1][~]{\text{\normalfont\AA}}

\renewcommand{\and}{\quad}

\newcommand{\pdbref}[1]{\href{http://www.rcsb.org/pdb/explore.do?structureId=#1}{PDB:#1}}
\newcommand{\arxiv}[2][UNDEFINED]{\href{https://arxiv.org/abs/#2}{\ifthenelse{\equal{#1}{UNDEFINED}}{arxiv.org/abs/#2}{#1}}}

\newcommand{\figref}[2][]{\hyperref[fig:#2]{Figure\@~\ref*{fig:#2}#1}}
\newcommand{\tabref}[1]{\hyperref[tab:#1]{Table \ref*{tab:#1}}}
\renewcommand{\eqref}[2][]{\hyperref[eq:#2]{Equation#1\@~\ref*{eq:#2}}}
\newcommand{\panelref}[2][]{%
    \ifthenelse{\boolean{onechapter}}{%
        \hyperref[panel:#2]{Panel\@~``\nameref{panel:#2}#1''}%
    }{%
        \hyperref[panel:#2]{Panel\@~\ref*{panel:#2}#1}%
    }%
}
\newcommand{\secref}[2][n]{%
    \hyperref[sec:#2]{%
        \ifthenelse{\equal{#1}{n} }{Section\@~\ref*{sec:#2}}{}% just number
        \ifthenelse{\equal{#1}{nn}}{Section\@~\ref*{sec:#2} ``\nameref{sec:#2}''}{}% nm & nr
        \ifthenelse{\equal{#1}{N} }{``\nameref{sec:#2}''}{}% just quoted name
        \ifthenelse{\equal{#1}{NN} }{\nameref{sec:#2}}{}% just name
    }%
}
\newcommand{\chref}[2][n]{%
    \ifthenelse{\boolean{onechapter}}{%
        \ifthenelse{\equal{#2}{ChPref}     }{\arxiv[Chapter ``\nameref*{ch:#2}'']{1801.09442}}{}%
        \ifthenelse{\equal{#2}{ChIntroPS}  }{\arxiv[Chapter ``\nameref*{ch:#2}'']{1801.09442}}{}%
        \ifthenelse{\equal{#2}{ChDetVal}   }{\arxiv[Chapter ``\nameref*{ch:#2}'']{2108.02706}}{}%
        \ifthenelse{\equal{#2}{ChStrucAli} }{\arxiv[Chapter ``\nameref*{ch:#2}'']{1801.09442}}{}%
        \ifthenelse{\equal{#2}{ChDBClass}  }{\arxiv[Chapter ``\nameref*{ch:#2}'']{1801.09442}}{}%
        \ifthenelse{\equal{#2}{ChFunc}     }{\arxiv[Chapter ``\nameref*{ch:#2}'']{1801.09442}}{}%
        \ifthenelse{\equal{#2}{ChIntroPred}}{\arxiv[Chapter ``\nameref*{ch:#2}'']{1712.00407}}{}%
        \ifthenelse{\equal{#2}{ChHomMod}   }{\arxiv[Chapter ``\nameref*{ch:#2}'']{1712.00425}}{}%
        \ifthenelse{\equal{#2}{ChSSPred}   }{\arxiv[Chapter ``\nameref*{ch:#2}'']{1801.09442}}{}%
        \ifthenelse{\equal{#2}{ChFuncPred} }{\arxiv[Chapter ``\nameref*{ch:#2}'']{1801.09442}}{}%
        \ifthenelse{\equal{#2}{ChIntroDyn} }{\arxiv[Chapter ``\nameref*{ch:#2}'']{1801.09442}}{}%
        \ifthenelse{\equal{#2}{ChThermo}   }{\arxiv[Chapter ``\nameref*{ch:#2}'']{1801.09442}}{}%
        \ifthenelse{\equal{#2}{ChMD}       }{\arxiv[Chapter ``\nameref*{ch:#2}'']{1801.09442}}{}%
        \ifthenelse{\equal{#2}{ChMC}       }{\arxiv[Chapter ``\nameref*{ch:#2}'']{1801.09442}}{}%
    }{
    \hyperref[ch:#2]{%
        \ifthenelse{\equal{#1}{n} }{Chapter \ref*{ch:#2}}{}% just number
        \ifthenelse{\equal{#1}{nn}}{Chapter \ref*{ch:#2} ``\nameref{ch:#2}''}{}% name & number
        \ifthenelse{\equal{#1}{N} }{``\nameref{ch:#2}''}{}% just name
      }%
  }%
}
\newcommand{\chrefname}[1]{\hyperref[ch:#1]{Chapter \ref*{ch:#1} ``\nameref{ch:#1}''}}
\newcommand{\partref}[1]{\hyperref[#1]{Part \ref*{#1}}}
\newcommand{\appref}[1]{\hyperref[app:#1]{Appendix \ref*{app:#1}}}

\newcommand{\figsource}[1]{\protect\footnote{Figure source location: \url{#1}}}

\newenvironment{cenum}[1][\itshape i)\upshape\ ]
{\begin{compactenum}[#1]} {\end{compactenum}}

\renewcommand{\arraystretch}{1.3}

\makeatletter \@beginparpenalty=5000 \makeatother

\newenvironment{panel}[1][]{
  \begin{figure}[htb]
    \begin{mdframed}[%
        outerlinewidth=0,%
        linecolor=CornflowerBlue!30,%
        backgroundcolor=CornflowerBlue!30,%
        innerleftmargin=14,%
        innerrightmargin=14,%
      ]
      \ifthenelse{\equal{#1}{}}{}{
        \stepcounter{panel}
		\subsection*{#1} 
      }
}{%
    \end{mdframed}
  \end{figure}
}

\newenvironment{bgreading}[1][]{
  \begin{mdframed}[%
      outerlinewidth=0,%
      linecolor=CornflowerBlue!30,%
      backgroundcolor=CornflowerBlue!30,%
      innerleftmargin=14,%
      innerrightmargin=14,%
    ]
	\ifthenelse{\equal{#1}{}}{}{
        \stepcounter{panel}
    	\subsection*{#1} 
    }
}{%
  \end{mdframed}
}

\usepackage{listings}
\definecolor{backcolour}{rgb}{0.95,0.95,0.92}
\definecolor{codegreen}{rgb}{0,0.6,0}
\definecolor{codegray}{rgb}{0.5,0.5,0.5}
\definecolor{codered}{rgb}{0.8,0,0.0}
\definecolor{codeblue}{rgb}{0.0,0,0.8}
\lstdefinestyle{codeStyle}{
    backgroundcolor=\color{backcolour},   
    commentstyle=\color{codegreen},
    keywordstyle=\color{codeblue},
    numberstyle=\tiny\color{codegray},
    stringstyle=\color{codegray},
    numbers=left,                    
    tabsize=2
} 
\newcommand{\sfrac}[2]{#1/#2}
\newcommand{\bfrac}[2]{#1/(#2)}

\makeindex

\begin{document}

\setboolean{onechapter}{true}

\pagestyle{fancy}
\lhead[\small\thepage]{\small\sf\nouppercase\rightmark}
\rhead[\small\sf\nouppercase\leftmark]{\small\thepage}
\newcommand{\innerfoot}{\footnotesize{\sf{\copyright} Feenstra \& Abeln}, 2014-2023}
\newcommand{\outerfoot}{\footnotesize \sf Intro Prot Struc Bioinf}
\lfoot[\outerfoot]{\innerfoot}
\cfoot{}
\rfoot[\innerfoot]{\outerfoot}
\renewcommand{\footrulewidth}{\headrulewidth}

\mainmatter
\setcounter{chapter}{12}
\chapterauthor{\JvG*~\JvGid \and \HM~\HMid \and \EvD~\EvDid \and \MD~\MDid \and \IH~\IHid \and \AG~\AGid \and \SA*~\SAid \and \AF*~\AFid}
\chapterfigure{\includegraphics[width=0.5\linewidth]{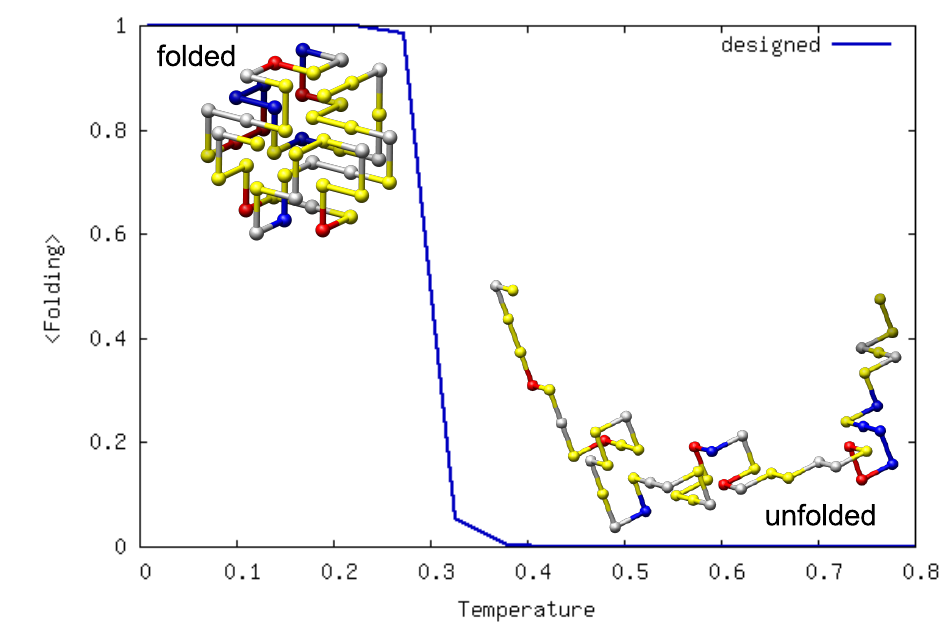}}
\chapterfootnote{* editorial responsability}
\chapter{Thermodynamics of Protein Folding}
\label{ch:ChThermo}

\ifthenelse{\boolean{onechapter}}{\tableofcontents\newpage}{}

In the previous chapter, \chref[N]{ChIntroDyn}, we introduced the concept of \emph{free energy} in the context of protein folding. The goal of that chapter was to start building an intuitive feeling of the meaning of the fundamental concepts that are important for this topic. Given a free energy landscape for the conformational space of a protein, we explained that conformations with low free energy are the most stable states of a protein (see \figref{ChIntroDyn-FreeEnergy}). In other words, these represent the states that are most probable, and proteins will typically spend most of the time in these low free energy states. The most stable state of a protein under normal conditions (e.g.,\@ room temperature and neutral pH) is often referred to as the native state of that protein. This is typically also the functional state of the protein. Although the protein is mostly in its native state under normal conditions, some fraction of the molecules may in fact be unfolded (typically as little as 1\% or even less).

In this chapter, we will take a step back and try to get a deeper, more formal understanding of the concept of free energy in terms of the \emph{entropy} and \emph{enthalpy}; to this end, we will first need to better define the meaning of equilibrium, entropy and enthalpy. When we understand these concepts, we will come back for a more quantitative explanation of protein folding and dynamics.

\section{Equilibrium and Dynamics}
\label{clasToStat}
In molecular simulations of proteins, we consider proteins in a \emph{dynamic equilibrium}. We do this under the assumption that over time, all systems move towards equilibrium. This means that the system will eventually reach a state where there is no net flow of energy or molecules between different parts of the system. 

\begin{figure}
  \centerline{
    \includegraphics[width=0.7\linewidth]{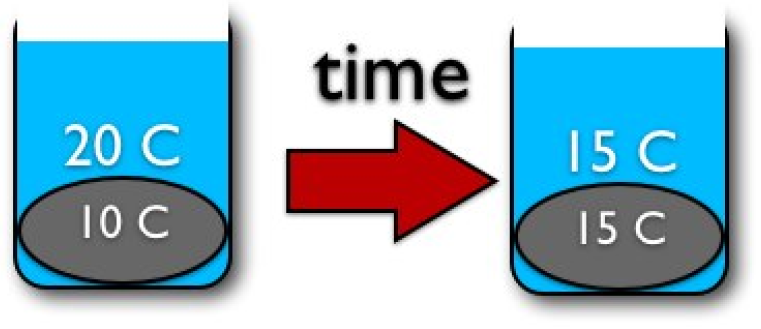}
  }
  \caption{Thermal Equilibrium. If a cold metal object ($T_1 = 10^\circ C$) is placed into a warm-water bath ($T_2 = 20^\circ C$), the two components will eventually exchange energy until they have reached the same temperature ($T\textsubscript{equilibrium} = 15^\circ C$).
}
 \label{fig:ChThermo:thermalEquilibrium}
\end{figure}

Take for example thermal equilibrium. When a cold metal object is placed into a warm water bath \figref{ChThermo:thermalEquilibrium}, the two subsystems will exchange energy in the form of heat. However, initially the warm water will have higher thermal energy than the metal object, and there will be larger energy flow from the water into the metal. Therefore, the water will (gradually) cool down and the metal object will become warmer. At a certain point, the temperature of the water and the metal object have become equal, and there will be no net energy flow. The system has now reached thermal equilibrium. Note that this is a macroscopic property of the system; although there is no net flow of energy in the water bath, at the microscopic level, individual atoms in the system can (and will) still exchange energy when they bump into one another. We will elaborate on macro- vs microscopic properties later in this chapter. 

Equilibrium also applies to protein folding and unfolding. When the system is in equilibrium, individual molecules can still (stochastically) switch between the folded and the unfolded state, but the total number of molecules in each state remains constant. In other words, although protein molecules are still switching between states, the probability to be in a certain state is constant. As highlighted before in \chref{ChIntroDyn}, the probability of being in a given state in the system is directly related to the free energy of that state: states with a low free energy have a high probability to be encountered, and are therefore considered to have high stability.

Finally, before we dive into the detailed material, we will provide a brief explanation of the difference between classical and statistical thermodynamics. We will need both, and understanding this difference will help you while reading the rest of the chapter. Classical thermodynamics is used to describe macroscopic properties of the system, such as the distribution of energy in the warm water bath described above, but does not consider microscopic fluctuations, such as the folding or unfolding of individual protein molecules. Statistical thermodynamics, also referred to as (equilibrium) statistical mechanics, describes how fluctuations at the microscopic (molecular) level of the system lead to the macroscopic behaviors that can be observed in experiments. Therefore, if we want to gain more mechanistic insight into properties of a process like protein folding, stability or function, we need to include the principles of statistical thermodynamics.

In the next sections, we begin by explaining the fundamental laws of thermodynamics, followed by the entropy and enthalpy. Finally, we will bring these concepts together to understand protein folding and dynamics from a thermodynamic perspective.

\section{Thermodynamic laws}
\label{sec:ThermoLaws}

The laws of thermodynamics were first formulated in the 19th and early 20th century. They have several fundamental implications for protein dynamics, and many checks, tricks and assumptions used in simulations are based on these laws. (Note that in literature the numbering of these rules is not always consistent.)

\begin{itemize}
\item[\textbf{Zeroth law of thermodynamics.}] The zeroth law of thermodynamics states that if we have three states (A, B and C) in a system, and both A \& B and B \& C are in equilibrium, then A \& C must also be in equilibrium. 

\item[\textbf{First law of thermodynamics.}] The first law is equivalent to the law of conservation of energy; energy cannot be created or destroyed, just transformed from one form to another. This can be rephrased by saying that the amount of total energy in the system does not change: 
\begin{equation}
\label{eq:ChThermo:firstlaw}
\frac{\partial E_{tot}}{\partial t} = 0
\end{equation}
\noindent
Examples of consequences of the first law: 
\begin{cenum}[1)]
\item Motor proteins in a cell use ATP to move along a microtubule \cite{Hirokawa1998, Mondal2017DriveCells}. This process converts internal chemical energy (in the form of ATP) into kinetic energy or work (moving forward with a certain velocity), and heat.
\item The F-ATPase found in bacteria and mitochondria can use ATP to move protons (H\textsuperscript{+}) across the cell membrane, or, reversibly, use a concentration gradient of protons to synthesize ATP \cite{Mitchell1961CouplingMechanism, Frasch2022F1FOMechanisms}. Here, chemical energy from the ATP molecule is converted into a different kind of chemical energy in the form of an electrochemical (charge \& pH) gradient across the cell membrane.
\item Friction converts kinetic energy (movement) into thermal energy (heat). For example, if you rub your hands together because you are feeling cold), they become warmer. 
\end{cenum}

\item[\textbf{Second law of thermodynamics.}] The second law of thermodynamics states that, for an isolated system, the entropy will never decrease (we will define entropy in the next section; for now, think of it as a measure for the amount of chaos/disorder). In other words, the entropy will keep increasing until it reaches the maximum possible value within the physical constraints of the system and then remain constant:
\begin{equation}
\frac{\partial S}{\partial t} \ge 0
\end{equation}
This also means that an isolated system always evolves to thermodynamic equilibrium (a state with no net flow of energy or molecules).

\noindent
Examples of consequences of the second law: 
\begin{cenum}[1)]
\item The gas atoms in a room will not spontaneously go to one corner in that room.
\item A purely hydrophilic peptide sequence will not spontaneously fold in a hydrophilic solution, as the folded state is entropically unfavorable.
\item Mixed red and blue marbles in a box will not separate spontaneously when shaking the box (we will use this example in \secref[nn]{ChThermo:Entropy}). 
\end{cenum}

\item[\textbf{Third law of thermodynamics.}] The entropy of a system at absolute zero (zero Kelvin), is constant and zero (actually close to zero due to quantum mechanical effects, and kinetically trapped states in so-called glassy systems) \cite{Kittel1980ThermalPhysics}. The third law has an interesting consequence: later in this chapter, we will see that entropy becomes unimportant near zero Kelvin (as the product of $T$ and $S$ approaches zero in \eqref{ChThermo:FETS}). 

\noindent
Examples of consequences of the third law: 
\begin{cenum}[1)]
\item There is no vibration of atoms in molecules at zero Kelvin (aside from quantum-mechanical effects).
\item At low temperature, a protein will optimize its internal interactions, i.e.,\@ minimise its total internal energy. (This property is the foundation of a simulation technique called `simulated annealing', as we will see later in \chref{ChMD} and \chref{ChMC}).
\item At high temperature, a protein will unfold, i.e.,\@ maximise its entropy.
\end{cenum}
\end{itemize}

Note that these laws apply to all physical systems, not just to the atoms, molecules and other details that we find important when investigating biological protein structures. Think of for example steam engines, the weather or galactic systems.

\section{Entropy}
\label{sec:ChThermo:Entropy}
In this section we will explain the concept of entropy in more detail. Entropy is a quantification of the amount of conformational freedom of the system. Loosely speaking, one may think of entropy as quantifying the amount of chaos or disorder in the system. More strictly, it is the number of possible microstates (structural conformations) that are accessible in a given macrostate (e.g., folded/unfolded state of a protein), as determined by the physical conditions.

\begin{figure}
  \centerline{
    \includegraphics[width=0.9\linewidth]{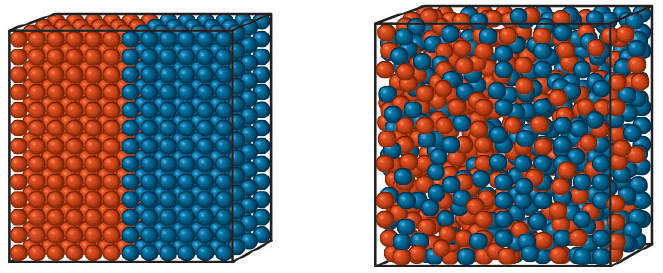}
  }
  \caption{Box with marbles. Initially, the marbles are sorted with the red marbles on one side and the blue marbles on the other side of the box (left panel). If the box is shaken, the marbles will move around randomly. In the equilibrium state, the marbles are distributed homogeneously over the box (right panel).}
\label{fig:ChThermo:marbles}
\end{figure}

To illustrate this concept on a simple system, take for example a box with spherical red and blue marbles that do not interact (i.e.,\@ there are no attractive or repulsive forces between the marbles). Initially, the marbles are sorted and all red marbles are at one side of the box and all blue marbles are at the other side (left panel \figref{ChThermo:marbles}). If you shake the box, the marbles will start moving around. After the box has been shaken for a sufficient amount of time, the red and blue marbles will be randomly distributed over the box. The number of red and blue marbles at each side of the box is now approximately equal. If you keep shaking the box, marbles will keep moving from one side to the other and back, but the total number of red and blue marbles at each side will no longer change. The system has reached equilibrium (right panel \figref{ChThermo:marbles}).

\begin{panel}[How to calculate the multiplicity of a state]
\label{panel:multiplicity}
The multiplicity of a state $A$ can be calculated by
\begin{equation}
\Omega_A = \frac{N!}{k_A! (N - k_A)!}
\end{equation}
where $N$ equals the total number of conformations and $k_A$ equals the number of conformations in one particular state. 

In our example of the marbles, imagine we have ten red and ten blue marbles in the box and that only ten marbles can fit into one side of the box (see \figref{ChThermo:marbles}). In this case, the number of conformations equals the total number of marbles at one side of the box, hence ${N=10}$. $k_A$ represents the number of red marbles at that side of the box. There are only two ways to have all red marbles at the same side, namely ${k_A=0}$ and ${k_A=10}$. However, there are 252 ways to distribute the marbles equally over the two sides of the box (if ${k_A=5}$, $\Omega_A = \frac{10!}{5!\cdot5!} = \frac{10\cdot9\cdot8\cdot7\cdot6\cdot5\cdot4\cdot3\cdot2\cdot1}{5\cdot4\cdot3\cdot2\cdot1\,\cdot\,5\cdot4\cdot3\cdot2\cdot1} = \frac{30240}{120} = 252$). Note that we do not care exactly which marbles are in which position, but only the total distribution of the differently coloured marbles over the sides of the box.
\end{panel}

Is it impossible for the red and blue marbles to spontaneously return to the initial (sorted) state? Technically, no. However, the probability of such an event is extremely low. The probability of being in a certain state is dependent on the number of ways to reach that state. This is called the multiplicity of a state, and is denoted by $\Omega$ (omega). For example, if four out of ten blue marbles are on the left side of the box, it does not matter which four these are, you will observe the same state, namely four blue marbles on the left side and six on the right side of the box. 

Another example is coin tossing. If you flip a coin three times and throw two heads and a tail, it does not matter if you throw two heads first and then a tail or first a tail and then two heads. Either way you reach the same state, namely two out of three heads. In the context of protein folding, the multiplicity would indicate the number of conformations in each state (e.g., folded/unfolded).   
The probability of each state can be calculated by dividing the multiplicity of that state by the sum of the multiplicities of all states in the system (i.e.,\@ the total number of ways to arrange the marbles in the box -- see \panelref{multiplicity}):

\begin{equation}
\label{eq:ChThermo:POmega}
{p_A} = \frac{\Omega_A}{\sum\limits_{X}{\Omega_X}}
\end{equation}
where ${p_A}$ equals the probability of state $A$, $\Omega_A$ is the multiplicity of state $A$ and $\sum\limits_{X}{\Omega_X}$ is the sum of the multiplicities of all states in the system.

From this equation one can see that in a system lacking attracting or opposing forces the largest multiplicity also has the largest probability. A large multiplicity implies a lot of conformational freedom, i.e.,\@ high entropy, which is directly related to the multiplicity $\Omega$ by:

\begin{equation}
\label{eq:ChThermo:SOmega}
S_A = {k_B} \ln{\Omega_A}
\end{equation}
where $S_A$ is the entropy of state $A$, ${k_B}$ is the Boltzmann constant (which relates temperature to energy per molecule \cite{Fischer2019}) and $\Omega_A$ is the multiplicity of state $A$. Thus, a state with a high multiplicity also has a high entropy, which makes that state more favorable. Note that in complex continuous systems like proteins in solution, the number of conformations is infinite and so we cannot count the absolute number of conformations. However, we can use the relation between multiplicity and probability to get an estimate of entropic differences between states from simulations (more on that later, in \secref[nn]{ChThermo:freeenergy}).

\section{Enthalpy}
\label{sec:ChThermo:enthalpy}
The enthalpy is the internal energy of the system plus the product of pressure and volume. The internal energy consists of kinetic energy (movement), thermal energy (heat), potential and interaction energies, amongst others. The more favourable interactions a molecule has, the lower the enthalpy. An example of interaction energy is the Van der Waals interaction, which can be described by the Lennard-Jones interaction potential (more detail on interaction potentials, including the Lennard-Jones in \chref{ChMD}). Other forms of interactions are polar interactions (e.g.,\@ hydrogen bonds), electrostatic interactions and hydrophobic interactions (e.g.,\@ $\pi$-stacking).

Note, that a favourable interaction reduces the enthalpy. The interaction energy at infinite distance is typically defined as zero; this means that a favorable interaction will have a negative energy.

\section{Free energy}
\label{sec:ChThermo:freeenergy}
So far, we have explained two thermodynamic concepts that affect the systems of interest: entropy and enthalpy. Now we will combine these two concepts to understand the stability of states in the system. As we have seen in \chref{ChIntroDyn}, the more stable a state is, the lower the free energy associated with that state is. If the entropy and enthalpy of the state are known, we can directly calculate the free energy of that state using

\begin{equation}
\label{eq:ChThermo:FETS}
F = H - TS
\end{equation}
where $H$ is the enthalpy, $T$ is the temperature in Kelvin and $S$ is the entropy. One can see from this formula that a macroscopic state with a high number of favorable interactions (low enthalpy) and/or low number of unfavorable interactions (which would increase the enthalpy), has a low free energy. Conversely, a state with fewer favorable interactions or more unfavourable interactions, will have a higher free energy. 

\begin{figure}
\centerline{
\includegraphics[width=1.05\linewidth]{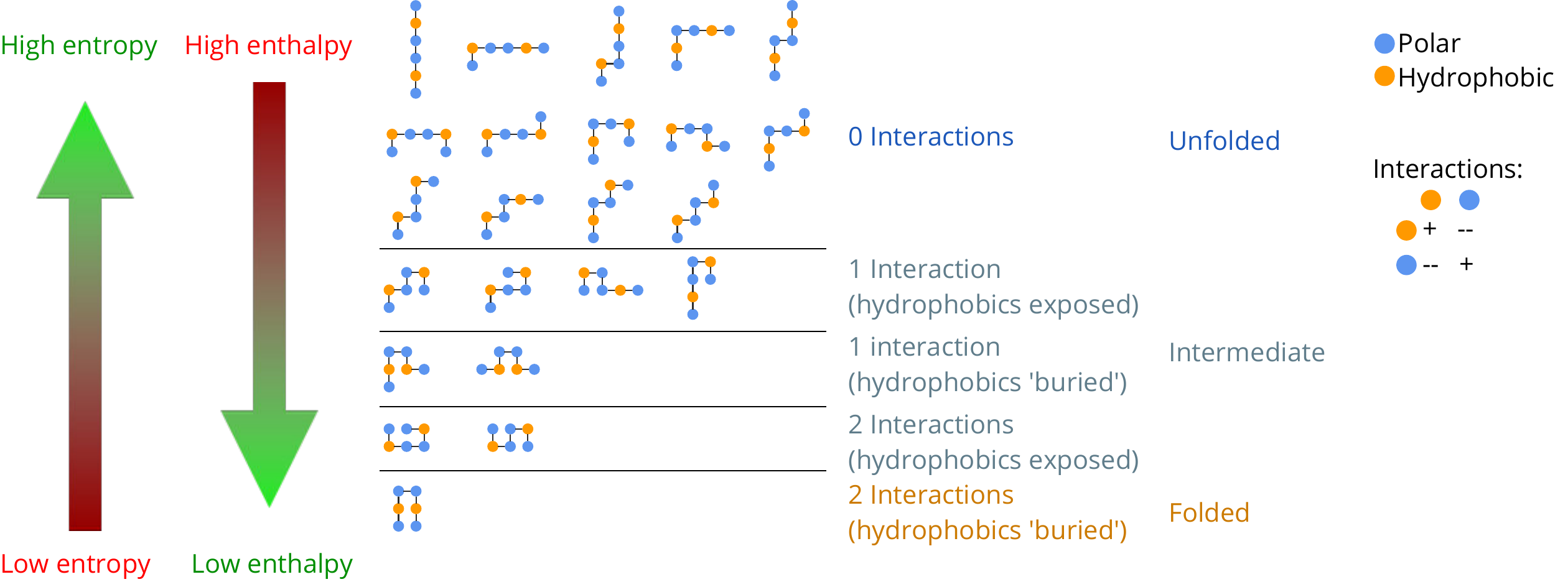}
}
\caption{Conformational enthalpy and entropy in an hydrophobic-polar (HP) lattice model of protein folding. Here, atoms can only be at the intersections of a two-dimensional square lattice. The lowest energy is shown at the bottom, which corresponds to the `folded' state. Energy here is counted as the number of interactions, where H-H and P-P are favorable and H-P is not. The top row has no interactions, the second row one P-P contact, the third row one H-H contact, and the bottom (native state) has one H-H plus one P-P contact. The number of `possible' conformations per energy level (state) decreases as well, going from top (unfolded) to bottom (native/folded).
Image adapted from Martin Gruebele,  University of Illinois, USA \cite{Ballew1996}.}

\label{fig:ChThermo:2Dlattice}
\end{figure}

Additionally, \eqref{ChThermo:FETS} shows that increasing the entropy of the system also reduces the free energy. Usually there is a trade-off between reducing the enthalpy and increasing the entropy of a system. Having many favorable interactions (low enthalpy) reduces the conformational freedom of the protein, which will lead to a lower entropy, which is entropically unfavourable. On the other hand, the state with the maximum entropy usually consists of unfolded conformations, where there are more interactions with the solvent. But for a protein this also means more interactions between hydrophobic residues and the solvent, which is enthalpically unfavourable. An example of this is shown in \figref{ChThermo:2Dlattice}. The figure shows a model of a six-residue peptide on a square lattice, where the orange circles indicate hydrophobic residues, and blue circles hydrophilic residues. There is one conformation with the lowest enthalpy (the highest number of favourable internal interactions), this is the folded state at the bottom in \figref{ChThermo:2Dlattice}. The highest entropy state is the unfolded state with fourteen conformations, which all have no internal interactions, shown at the top of \figref{ChThermo:2Dlattice}. Depending on the conditions of the system (e.g.,\@ temperature, pH, other solvents), the folded, unfolded or one of the intermediate states will be the most stable.

Now that we know the relation between free energy, entropy and enthalpy (\eqref{ChThermo:FETS}), the next question is then, how do we determine these properties from a system. As explained in \secref{ChThermo:Entropy}, the entropy $S_A$ of a state A is directly related to its multiplicity via \eqref{ChThermo:SOmega}. The total energy of the state $E_A$ can be calculated from the weighted sum of the energies of the conformations in state A:
\begin{equation}
\label{eq:ChThermo:E_i}
E_A = \sum\limits_{i\in A}{E_i p_i}
\end{equation}

However, the calculation of the absolute values of the free energy contributions is far too complex for continuous systems such as proteins in solution, because the number of conformations to be considered is infinite. Nevertheless, the probability of a macroscopic state, such as the folded or unfolded state of a protein, can be approximated from simulations by determining the fraction of time spent in that state (see \panelref{ChThermo:derivation_F} to see how we get here using statistical thermodynamics):

\begin{equation}
\label{eq:ChThermo:flnp}
F_A \propto -k_B T \ln{(p_A)}
\end{equation}
where $F_A$ is the free energy of state $A$, $k_B$ is the Boltzmann constant, $T$ is the temperature in Kelvin and $p_A$ is the probability of the system to be in state $A$. Note that $F_A$ is proportional, not equal, to $p_A$. Thus, we cannot calculate the absolute value of the free energy of state $A$ using this approach. In practice, this is not a problem, since we are usually only interested in the \emph{free energy difference} between two states (e.g folded vs unfolded) to see which one is more favourable, rather than the absolute free energy of an individual state. This difference can be obtained using:
\begin{eqnarray}
\label{eq:ChThermo:FRel}
\Delta F_{A \rightarrow B} &=& F_B - F_A \nonumber\\
         &=& -k_B T \ln{(p_B)} - \left( -k_B T \ln{(p_A)} \right)
\end{eqnarray}
Thus, the free energy difference between two states equals:

\begin{equation}
\label{eq:ChTermo:pa_pb}
	 \Delta F_{A \rightarrow B} = -k_B T \ln{\left(\frac{p_B}{p_A}\right)}
\end{equation}
\eqref{ChTermo:pa_pb} is very important, since it tells us that the only thing we need to know to calculate the free energy difference between two states, is the relative amount of time proteins spend in each state; $p_A$ and $p_B$. As we will see in \chref{ChMD} and \chref{ChMC}, these probabilities can easily be determined from simulations (provided all states are sampled sufficiently).

\begin{bgreading}[Derivation of free energy using statistical thermodynamics] 

\label{panel:ChThermo:derivation_F}
Combining \eqref{ChThermo:SOmega}, \eqref{ChThermo:FETS}, and \eqref{ChThermo:E_i}, the formula for the free energy of state $A$ ($F_A$) becomes 

\begin{equation}
\label{eq:ChThermo:fOmega}
F_A = E_A - T S_A = \sum_{i}{E_i p_i} - k_B T \ln{\left( \Omega_A \right)}
\end{equation}
where $E_i$ is the energy of a conformation in the ensemble of conformations belonging to state A, $p_i$ is the probability of that conformation, $k_B$ is the Boltzmann constant, $T$ is the temperature in Kelvin and $\Omega$ is the multiplicity of the state. In a complex system such as a protein in solution, the multiplicity can usually not be calculated explicitly. However, in \panelref{multiplicity}, we explained that $\Omega_A = \bfrac{N!}{k_A!\left(N-k_A\right)}$. Applying this to \eqref{ChThermo:SOmega} ($S_A = {k_B} \ln{\Omega_A}$), using Stirling's approximation, which says that the logarithm of the factorial of a very large number can be approximated as $\ln (N!) \approx N \ln (N) - N$, \cite{Glazer2002} and plugging the results into \eqref{ChThermo:fOmega}, eventually yields:
\begin{equation}
F_A = \sum_{i}{E_i p_i} + k_B T \sum_{i}{p_i \ln{(p_i)}}
\label{eq:ChThermo:FETS2}
\end{equation}

\eqref{ChThermo:POmega} shows how to calculate the probability of a state if there are no interactions between the particles in the system (i.e.,\@ no enthalpy, so the system is entropy-driven), such as the example with the marbles in \secref{ChThermo:Entropy}. In a system where the particles do interact, such as a protein in solution, the probability of a conformation also depends on the energy of that conformation. This is captured in the Boltzmann distribution, which describes the probability $p_i$ of a state $i$ as function of its energy $E_i$ in an equilibrium situation:
\begin{equation}
\label{eq:ChThermo:Boltzmann}
p_i = \frac{e^{- \sfrac{E_i}{k_B T}}}{\sum_{i}{e^{- \sfrac{E_i}{k_B T}}}}
\end{equation}
where $p_i$ is the probability of a conformation, $E_i$ is the energy of that conformation, $k_B$ is the Boltzmann constant and $T$ is the temperature (in Kelvin). The denominator is called the \emph{Partition function}, denoted as $Z$: 
\begin{equation}
\label{eq:ChThermo:partition}
    Z ~ = ~ \sum_{i}{e^{- \frac{E_i}{k_B T}}}
\end{equation}
which simplifies \eqref{ChThermo:Boltzmann} to:
\begin{equation}
\label{eq:ChThermo:Boltzmann2}
p_i = \frac1Z{e^{- \frac{E_i}{k_B T}}}
\end{equation}
Substituting \eqref{ChThermo:Boltzmann2} back into \eqref{ChThermo:FETS2} and simplifying the result gives the total free energy $F$ of the system:
\begin{equation}
\label{eq:ChThermo:flnZ}
F_A = -k_B T \ln{(Z)}
\end{equation}
We can now directly relate the free energy to the probability of being in a certain state. \eqref{ChThermo:partition} defines the partition $Z$ as the sum of all Boltzmann factors of the conformations in the system. For a system like a solution of protein molecules, we can think of this as describing the how likely it is to find one molecule in a particular microstate at any given moment in time, or in other words, how the system is divided, \emph{partitioned}, across the different microstates. Then, we can reverse \eqref{ChThermo:Boltzmann2} to obtain  the probability $p_i$ of being in that state $i$, then 

\begin{equation}
F_i \propto -k_B T \ln{(p_i)}
\end{equation}
This probability can, at least in principle, be obtained from simulations or experiments. We will show examples of this, from simulation, in the next section, \secref[N]{ChThermo:free-energy-temperature}, and more in \chref[nn]{ChMD} and \chref[nn]{ChMC}.

\end{bgreading}

\subsection{Temperature Dependence of Free Energy Landscapes}
\label{sec:ChThermo:free-energy-temperature}

\begin{figure}
\centerline{\includegraphics[width=1.1\linewidth]{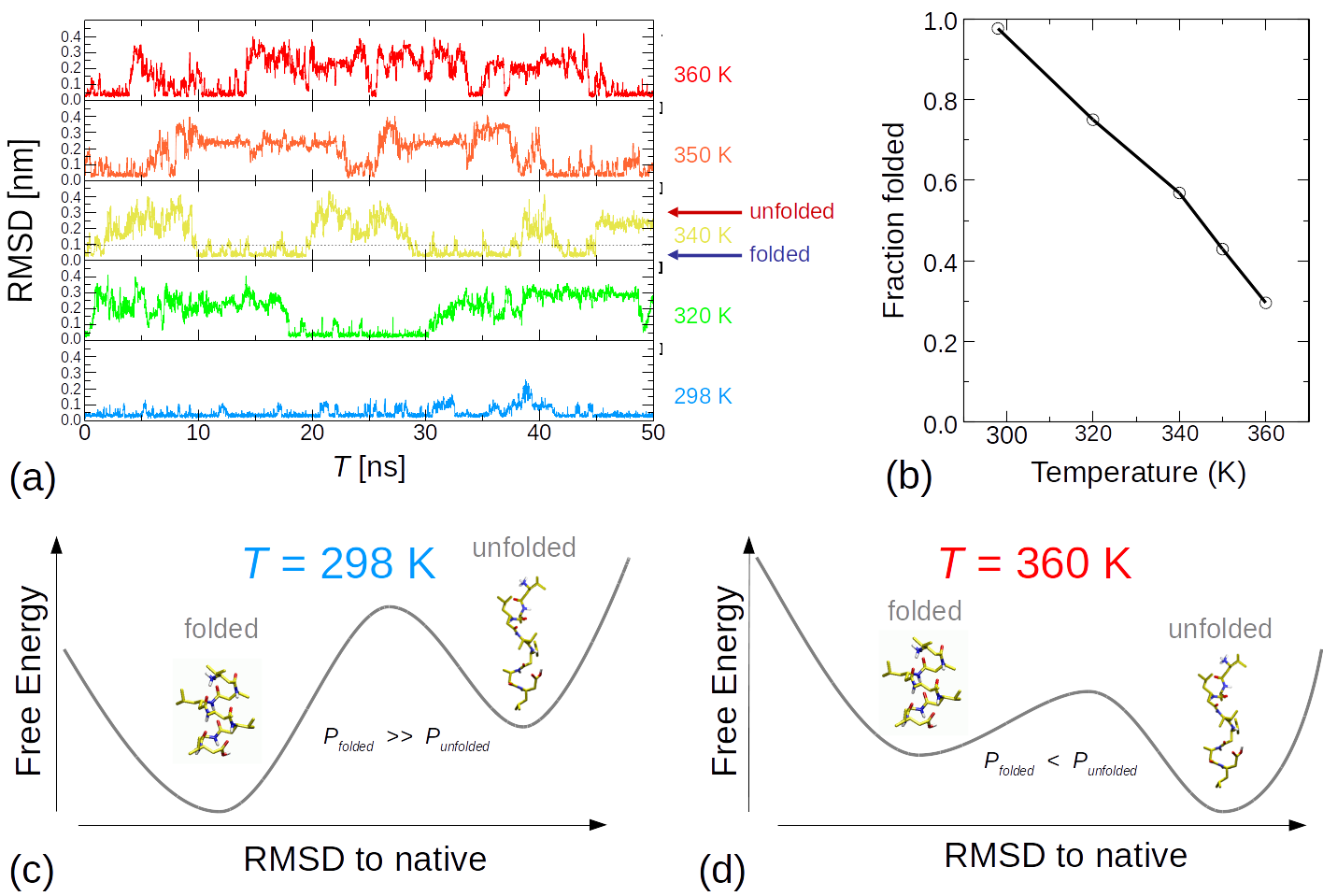}}
\caption{Temperature dependence of protein stability. (a) RMSD vs. time for 5 different temperatures: 298, 320, 340, 350 and 360 K. (b) Fraction folded as function of temperature, derived from the simulations shown in (a). At room temperature the protein is almost always in its folded state. As the temperature increases the protein is more in its unfolded state. (c) Schematic Free Energy diagrams corresponding to the lowest temperature (298K). The reaction coordinate used is the same as for (a): the RMSD to native. The folded state (left) has a lower RMSD, while for the unfolded state (right) it is high. The free energy of the folded state is lower, which indicates it is more stable than the unfolded state at this temperature. The barrier between folded and unfolded states limits the rate at which folding and unfolding events may happen. (d) Same, but for the highest temperature (360K). Now, the free energy of the unfolded state is lower, which indicates it is more stable than the folded state at 360K. The barrier between folded and unfolded states is somewhat lower, reflecting the higher rate at which folding and unfolding can be observed in panel (a). Panel (a), data for panel (b) and structures in (c) and (d) with permission from Daura \& Oostenbrink \cite{Daura1998}.}
\label{fig:ChThermo:reversible-peptide-temperature}
\end{figure}

One of the reasons we are interested in the thermodynamics of protein folding is so we can understand the temperature dependence of the stability. Proteins can fold reversibly, as we know from \citet{Anfinsen1973}. \figref[a]{ChThermo:reversible-peptide-temperature} shows simulations of a small (7 amino acids) peptide to illustrate this phenomenon. Each of the colors in the graph represents a time trace of the RMSD to the native structure over time for simulations at a different temperature. One can see in the figure that throughout the simulations, the RMSD increases and decreases, indicating that the protein unfolds and refolds several times. It is clear that as the temperature increases the protein spends more time in the unfolded state (high RMSD in \figref[a]{ChThermo:reversible-peptide-temperature}) which can be quantified by the probability (time spent) of the folded state as a function of temperature (\figref[b]{ChThermo:reversible-peptide-temperature}).

Let us see if we can understand this from a thermodynamic perspective. Recall, from \eqref{ChThermo:FETS}, that $\Delta F = \Delta H - T \Delta S$. In this example, we are talking about the free energy of folding, so the $\Delta$ here indicates the difference between the folded state and the unfolded state. A negative value of $\Delta F$ indicates the unfolded state is more favourable, whereas a positive value for $\Delta F$ indicates the folded state is more favourable. For the first term of the equation, the unfolded state has fewer favorable internal contacts than the folded state, so the unfolded state has a higher enthalpy: $\Delta H = H_{unfolded} - H_{folded} > 0$. For the second term of the equation, the unfolded state has more conformational freedom and thus a higher entropy than the folded state: $\Delta S = S_{unfolded} - S_{folded} > 0$. Since temperature in Kelvin ($T$) is always positive, in the formula $\Delta F = \Delta H - T \Delta S$, both $\Delta H$ and $T\Delta S$ are positive.

At low $T$, the absolute value of $T\Delta S$ will be small. Therefore, the favorable enthalpy of the folded protein structure to $\Delta F$ will outweigh the unfavourable entropy of the folded state at low temperatures, and the protein will spend most time in the folded conformation, as shown in \figref[c]{ChThermo:reversible-peptide-temperature}.

As $T$ increases, the absolute value of $T\Delta S$ will increase as well. At a sufficiently high value of $T$, the favorable enthalpy of the folded protein structure to $\Delta F$ will no longer outweigh the unfavourable entropy of the folded state, and at higher temperatures, the protein will spend most time in the unfolded conformation, as shown in \figref[d]{ChThermo:reversible-peptide-temperature}. This decrease in stability of the folded state at higher temperatures is exactly what is observed in \figref[a and b]{ChThermo:reversible-peptide-temperature}: as temperature increases, the protein spends relatively more time in the unfolded state because the unfolded state becomes more stable than the folded state. Thus, the importance of entropy increases with increasing temperature.

\begin{figure}
\centerline{
\includegraphics[width=0.7\linewidth]{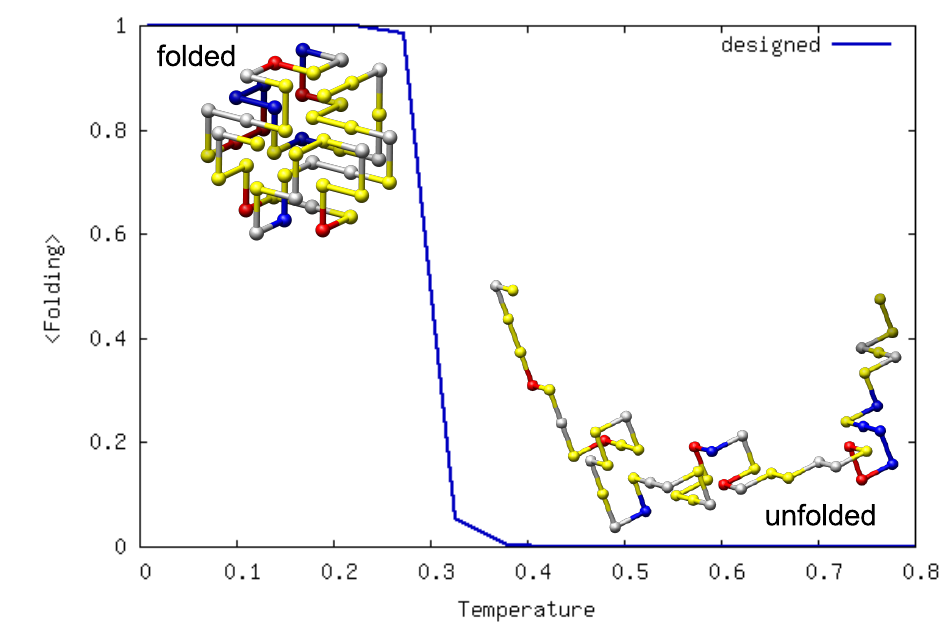}
}
\caption{Temperature dependence of protein folding. The horizontal axis shows temperature (the results are from a simplified lattice model of protein folding with reduced units for temperature). The vertical axis shows the extent of folding, 1 meaning fully folded and 0 unfolded (measured by the fraction of native contacts formed). At low temperature (T$<$0.2), the native state is stable and therefore the protein is folded. This folding is driven by the energetically favourable conformation where hydrophobic residues (yellow) are `shielded' in the interior of the protein structure. At high temperature (T$>$0.4), entropic effects win out over the energetic effects. This makes the unfolded state more stable. The unfolded state, naturally, has a higher entropy than the folded state, but has to pay the energetic cost of exposing hydrophobic residues to the water \cite{vanDijk2015,vanDijk2016}.}
\label{fig:ChThermo-lattice-folding-temperature}
\end{figure}

\figref{ChThermo-lattice-folding-temperature} shows another example of temperature dependence of protein folding, but here a 3D lattice model is used. Despite its simplicity, we can observe the transition from a stable folded state at low temperature, through a transition temperature at which both states are stable to some extent, to high temperatures where the unfolded state is the most stable \cite{vanDijk2015}. Although the model is very simple, it is realistic in the sense that such transition temperature effects are also observed for real proteins \cite{Thiriot2005StructuralBacteriophage}.

\section{From Microstates to Macrostates}
So far, we have discussed the formal definitions of the free energy, enthalpy and energy of a system and how they are related. An important distinction that has been mentioned, but not yet emphasized, is the distinction between \emph{microstates} and \emph{macrostates}. Microstates are individual conformations (in the previous sections of this chapter indicated with the subscript $i$), whereas macrostates describe the behavior of ensembles of conformations with very similar properties (indicated with the subscripts $A$ and $B$). For example, the folded and unfolded states with local energy minima in \figref[c,d]{ChThermo:reversible-peptide-temperature} are macrostates. The probability of a conformation calculated using the Boltzmann equation is an example of the probability of a microstate. Another example can be found in \figref[a]{ChIntroDyn-ensemble}, where each individual conformation shown of the structure is a microstate, and the collection of conformations is a macrostate. In the next two subsections we will explain two concepts that are important to understand how to relate microstates to macrostates, namely \emph{order parameters} and \emph{ensemble averages}.

\subsection{Order Parameters}
In a free energy profile (such as shown in \figref[c,d]{ChThermo:reversible-peptide-temperature}) the free energy is determined for different values of an \emph{order parameter}. An order parameter is a quantitative measure that can distinguish the relevant states of a system (also see panel \panelref{ChMD:Analysis}). For protein folding, frequently used order parameters are for example the number of native contacts (e.g.,\@ the number of internal hydrogen bonds), the RMSD to the native structure or the radius of gyration (a measure of the diameter of a flexible molecule, which indicates how compact or extended a protein conformation is). Order parameters are used to define the macrostates in the system. For example, we can define an RMSD threshold to determine whether a protein is folded or unfolded.

Choosing a suitable order parameter is very important when setting up a research project involving simulations. Not every order parameter is able to help answer every research question. For example, say that we are interested in the folded and unfolded state of a protein with a large disordered loop region  (e.g., a protein with two domains separated by a linker or some transmembrane proteins). During simulation, this region would be relatively flexible and move around a lot. As a result, the RSMD of this structure would be relatively high in both the folded and the unfolded state. Thus, the RMSD might not be able to distinguish the states of interest that well, and would not be a suitable order parameter. The number of native contacts on the other hand, would be less sensitive to the dynamics of the loop region(s), and might be more suitable in this case.

For the simple two-state examples of protein folding considered so far, a single order parameter suffices, but in many cases two or several more order parameters are used to establish a multi-dimensional free energy landscape. We will discuss an example of this in \chref[nn]{ChMC}, where we use native \textsl{vs.}\@ non-native contacts to describe an two-dimensional free energy landscape (see \secref[nn]{ChMC:lattice}).

\subsection{Ensemble Average}
\label{sec:ChThermo:EnsAvg}
Apart from free energies, we may also be interested in other properties that are characteristic of the system (e.g., enthalpy, radius of gyration, secondary structure content); Especially properties that can be measured experimentally, because those show that the simulations can reproduce behavior observed \textit{in vitro} or \textit{in vivo}, while providing mechanistic insights into the observed processes.

Because the values of the property of interest may not equal for all conformations within a state, and the probabilities of observing each conformation within a state may not be equal, we calculate a weighted average to describe that property for a state. This weighted average is called the \emph{ensemble average}. The ensemble average for a property $X$ in state $A$ is calculated by:

\begin{equation}
\label{eq:ChThermo:EnsAvg}
\langle X \rangle_A = \frac{\sum_{i \in A}{X_i p_i}}{\sum_{i \in A}{p_i}}
\end{equation}
where $X_i$ is the value of the property of interest $X$ in conformation $i$ and $p_i$ is the probability of conformation $i$. $i \in A$ is a notation indicating all conformations $i$ in state $A$. Note that $i$ here denotes microstates, and $A$ the (observable) macrostate, as introduced at the start of this section. From our simulation, we can directly calculate an ensemble average, by averaging a certain property $X$ over all the sampled conformations.

One example of the application of ensemble averages is the calculation of the enthalpy of a state. The enthalpy of a state is the ensemble average of the internal energies of all the conformations in the ensemble of that state:
\begin{equation}
\label{eq:ChThermo:EnsH}
H_A = \langle E \rangle_A = \frac{\sum_{i \in A}{E_i p_i}}{\sum_{i \in A}{p_i}}
\end{equation}
where $H$ is the enthalpy and $E$ is the internal energy (more on the relation between the two in the next section). To calculate the difference in enthalpy we can subtract the ensemble averages of the energies of the two states:
\begin{equation}
\label{eq:ChThermo:DeltaH}
\Delta H_{A,B} = \langle E \rangle_B - \langle E \rangle_A
\end{equation}
Thus, ensemble averages can help us quantify structural properties related to macrostates.

\begin{figure}[b]
  \centerline{
    \includegraphics[width=\linewidth]{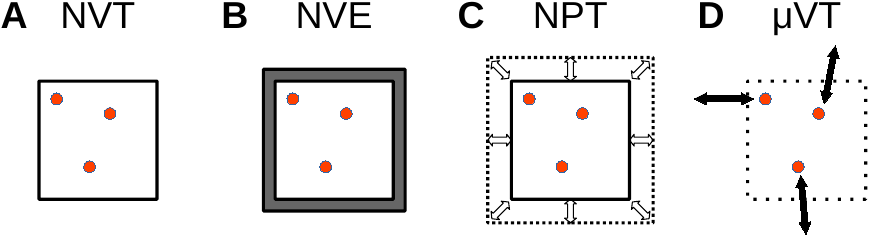}
  }
  \caption{A schematic representation of different ensembles. A: NVT ensemble, B: NPT ensemble, C: $\mu$VT ensemble, D: NVE ensemble.
  }
\label{fig:ChThermo-statensembles}
\end{figure}

\section{Ensembles}
\label{sec:ChThermo:Ensembles}

In molecular simulations, we generally consider the molecules of interest in an isolated environment (e.g., protein in a box of water with constant temperature and volume). We will not go into detail here, but generally this helps significantly simplify the calculations needed during the simulations. It is important to realise that thermodynamic relations are slightly different depending on the physical conditions chosen for the system. This gives rise to what is known in thermodynamics as different ensembles (not to be confused with ensemble averages), each with their own specific definition of free energy. In this section, we will (briefly) discuss several of these ensembles, as shown schematically in \figref{ChThermo-statensembles}. Each of them are defined by the three parameters that are constant under the specific conditions:
\begin{cenum}
\item[\textbf{NVT}:] Constant number of particles ($N$), volume ($V$) and temperature ($T$), typically encountered in Monte Carlo simulations (\chref{ChMC}). Known as the ``canonical'' ensemble.
\item[\textbf{NVE}:] The ``microcanonical ensemble'' is the natural situation in molecular dynamics simulations (\chref{ChMD}), where we typically define a fixed system ($N$), in a fixed size environment ($V$) and no exchange of energy allowed ($E$).
\item[\textbf{NPT}:] Instead of volume, here the pressure ($P$) is constant, as well as the number of particles ($N$) and temperature ($T$). This is closer to laboratory conditions, and still convenient for simulations. Therefore, in MD simulations, usually a thermostat and barostat are applied to the system such that the simulations are performed in NPT rather than NVE (\chref{ChMD}).
\item[\textbf{$\mu$VT}:] Instead of number of particles, here the chemical potential ($\mu$) is constant, which presumes exchange of particles with the surroundings. Additionally, the volume ($V$) and temperature ($T$) are constant. This so-called ``grand canonical'' ensemble is close to typical laboratory conditions, but hard to achieve in simulations.

\end{cenum}%
Note, that we do not go into why there are three parameters, and why they occur in these combinations (although there are more possible). For this, please refer to \citet{Schroeder}.

We will consider two ensembles, the NVT and NPT, in some more detail, as these are common in simulations. In the NVT ensemble, the number of particles $N$, the volume $V$ and the temperature $T$ are kept constant. This corresponds to the so called Helmholtz ensemble or canonical ensemble, and the Helmholtz free energy; which is what we defined in \eqref{ChThermo:FETS}, and is often written as:
\begin{equation}\label{eq:ChThermo:dGdHTdS}
    \Delta F = \Delta E - T\Delta S
\end{equation}

Here, $F$ is the Helmholtz free energy. $E$ is the internal energy in the system.  We typically cannot calculate or measure absolute values for energies, but only differences between states. This is not a problem since these relative free energies between states determines their stability, and therefore, the difference is the variable that needs to be determined. In protein folding for instance, these can be between the folded and unfolded states (e.g.,\@ \figref{ChIntroDyn-FreeEnergy}), or for the case of protein interactions, between the bound and unbound states.

In the NPT ensemble, the number of molecules $N$, pressure $P$ and temperature $T$ are constant, and the Gibbs free energy $G$ is minimized. The Gibbs free energy is defined as
\begin{equation}
	\Delta G = \Delta E - T\Delta S + PV = \Delta H - T\Delta S
\end{equation}
The Gibbs free energy is the Helmholtz free energy plus the product of the pressure and volume. The term $E + PV$ is called the enthalpy and is typically denoted as $H$. Due to the low compressibility of water, the $PV$ term can often be neglected in the systems we are interested in. Therefore, we sometimes consider the free energy without specifying whether we use the Helmholtz or the Gibbs ensemble.

\section{Conclusion}
In this chapter we have elaborated on the concepts introduced in \chref{ChIntroDyn} in a more formal way using principles from thermodynamics and statistical mechanics. We have mainly focused on how the free energy is related to the entropy and enthalpy, and how this affects the probability that a system is in a certain state. We have also illustrated how changes in the temperature affect the free energy landscape. Finally, we highlighted some key differences between microstates (individual conformations) and macrostates (ensembles of conformations with similar properties). In the next chapters, we will go into more depth on how simulation methods can be applied to study free energy landscapes.

\section*{Key concepts}
\begin{compactitem}
\item Free energy $F$ is a function of the internal energy $E$, entropy $S$ and temperature $T$: $F = E - TS$
\item Boltzmann's relation gives us the probability $p_i$ of conformation $i$ as function of its energy $E_i$: $p_i = \frac1Z{e^{\sfrac{-E_i}{k_B T}}}$ 
\item The free energy of a state is proportional to the probability of encountering that state: $F_A \propto -k_B T \ln (p_A)$
\item The free energy difference between two states can be calculated by: $\Delta F_{A \rightarrow B} = -k_B T \ln{\left( \sfrac{p_B}{p_A}\right)}$
\item Ensemble averages can be used to describe the general behaviour of the microstates belonging to a macrostate
\item Different thermodynamic ensembles are used to describe the physical properties of the system of interest
\item An order parameter is needed to distinguish different states of the system; only then does it become possible to calculate free energies
\end{compactitem}

\section*{Further Reading}
\label{sec:ChThermo:reading}
\begin{compactitem}
\item ``Physical Biology of the Cell'' -- \citet{Phillips2012}
\item ``Statistical Mechanics - A survival guide'' -- \citet{Glazer2002}
\item ``The Real Reason Why Oil and Water Don't Mix'' -- \citet{Silverstein1998}
\item ``An Introduction to Thermal Physics'' -- \citet{Schroeder}
\end{compactitem}

\section*{Author contributions}
{\renewcommand{\arraystretch}{1}
\begin{tabular}{@{}ll}
\ACtxt: &   JvG, EvD, HM, KAF, SA \\
\ACfig: &   JvG, AG, KAF, SA \\
\ACref: &   JvG, JV, IH, KAF, SA\\
\ACproof:&  JvG, HM, KAF, JV, SA \\
\ACfb:  &   JB, JG \\
\ACeds: &  JvG, KAF, SA
\end{tabular}}

\mychapbib

\clearpage

\cleardoublepage

\end{document}